\documentclass[a4paper,fleqn,usenatbib]{mnras}

\usepackage{newtxtext,newtxmath}
\usepackage[T1]{fontenc}
\usepackage{ae,aecompl}

\usepackage{graphicx}	% Including figure files
\usepackage{amsmath}	% Advanced maths commands
\usepackage{amssymb}	% Extra maths symbols
\usepackage{enumitem}
\setlist{nosep} 
\usepackage{xspace}
\usepackage{bm}
\usepackage{multirow}
\usepackage{listings}
\newcommand{\Enzo}{\textsc{Enzo}\xspace}
\newcommand{\yt}{\textsc{yt}\xspace}
\newcommand{\Music}{\textsc{MUSIC}\xspace}
\newcommand{\Rockstar}{\textsc{Rockstar}\xspace}

\newcommand{\flt}[1]{\overline{#1}}
\newcommand{\fav}[1]{\widetilde{#1}}
\newcommand{\EMF}{\bm{\mathcal{E}}}
\newcommand{\tu}[1][]{\tau_{ij}^{\mathrm{u} #1}}
\newcommand{\tb}[1][]{\tau_{ij}^{\mathrm{b} #1}}
\newcommand{\bra}[1]{\left( #1 \right)}
\newcommand{\Ekinsgs}{E^{\mathrm{u}}_{\mathrm{sgs}}}
\newcommand{\Emagsgs}{E^{\mathrm{b}}_{\mathrm{sgs}}}
\newcommand{\Msun}{M$_{\sun}$\xspace}
\newcommand{\Ma}{\mathrm{M_A}}
\newcommand{\Ms}{\mathrm{M_S}}

\title[Turbulent, magnetised SMBH seed formation]{Intermittent fragmentation and 
statistical variations during gas collapse in magnetised atomic cooling haloes}

\author[P. Grete et al.]{
P. Grete,$^{1}$\thanks{E-mail: grete@pa.msu.edu}
M. A. Latif,$^{2}$
D. R. G. Schleicher,$^{3}$
and W. Schmidt$^{4}$
\\
$^{1}$
Department of Physics and Astronomy, Michigan State University, East Lansing, MI 48824, USA
\\
$^{2}$
Physics Department, College of Science, United Arab Emirates University, PO Box 15551, Al-Ain, UAE
\\
$^{3}$
Departamento de Astronom\'ia, Facultad Ciencias F\'isicas y Matem\'aticas,
Universidad de Concepci\'on, Av. Esteban Iturra s/n Barrio Universitario, Casilla 160-C, Chile
\\
$^{4}$
Hamburger Sternwarte, Universität Hamburg, Gojenbergsweg 112, D-21029 Hamburg, Germany
}

\date{Accepted XXX. Received YYY; in original form ZZZ}

\pubyear{2019}

\begin{document}
\label{firstpage}
\pagerange{\pageref{firstpage}--\pageref{lastpage}}
\maketitle

% Abstract of the paper
\begin{abstract}
Observations reveal the presence of supermassive black holes (SMBH) as early as ${\sim}700$~million years
after the Big Bang.
Their formation path is still subject to current debate.
We explore the influence of magnetic fields, which are strongly amplified via the turbulent 
small-scale dynamo, on the formation of SMBH seeds within the direct collapse scenario.
In this study, we perform for the first time cosmological magnetohydrodynamic large eddy
simulations that employ a model for unresolved, compressible MHD turbulence. 
In total we perform 36 simulations for 9 haloes each with two different initial magnetic field strengths, and with
and without employing the unresolved turbulence model.
We make use of the adaptive mesh refinement approach to achieve an effective spatial resolution 
of less than one proper astronomical unit.
We consider a regime where cooling is regulated by atomic hydrogen and the molecular 
hydrogen gets dissociated by a strong radiation field.
Our main finding is that the majority of the gas properties in the haloes at the final output 
are predominantly determined by the run-away gravitational collapse.
Turbulence is supersonic and super-Alfv\'enic in all cases, and magnetic fields are amplified to
an approximately dynamically relevant regime.
Finally, fragmentation during the collapse is intermittent and mass accretion rates 
range from $0.2-3$\,\Msun/yr.
This suggests that the presence of strongly amplified magnetic fields and turbulence provides additional pressure support on small scales and make the direct collapse a viable scenario for the formation of massive objects under the required ambient conditions.
\end{abstract}

% Select between one and six entries from the list of approved keywords.
% Don't make up new ones.
\begin{keywords}
(magnetohydrodynamics) MHD --
turbulence --
methods: numerical --
quasars: supermassive black holes --
early Universe --
cosmology: theory
\end{keywords}

%%%%%%%%%%%%%%%%%%%%%%%%%%%%%%%%%%%%%%%%%%%%%%%%%%

%%%%%%%%%%%%%%%%% BODY OF PAPER %%%%%%%%%%%%%%%%%%

\section{Introduction}
Supermassive black holes (SMBH) with up to a billion solar masses are common in the centres of 
present-day galaxies \citep{Kormendy95,Tremaine02}, and their presence has 
been confirmed as early as ${\sim}700$ million years after the 
Big Bang \citep{Fan2003,Mortlock2011,Wu2015,2018arXiv180706055S,Banados2018}. 
The formation of such massive black holes~(BH) is still an enigma. 
It is important to obtain a physical understanding how they formed and how efficiently 
they grew in the first billion years. 
Different pathways to form supermassive black holes have been 
suggested \citep{1984ARA&A..22..471R,Volonteri2010, LatifFerrara2016}. 
There are three main mechanisms that can potentially lead to the formation of SMBHs. 
They include the accretion and merging of Pop III remnants 
\citep{2002ApJ...567..532H,2012ApJ...756L..19W,2001ApJ...551L..27M}, 
the relativistic instability in dense stellar cluster 
\citep{2010MNRAS.409.1057D,2012MNRAS.421.1465D,2014MNRAS.442.3616L,2015MNRAS.451.2352K,
2018MNRAS.476..366B,2018A&A...614A..14R,2018arXiv181202052S,10.1093/mnras/stz315} 
as well as the direct collapse of protogalactic gas clouds
\citep{Bromm2003,2006MNRAS.371.1813L,Wise2008,Latif2013BHform,2015A&A...578A.118L,
2015MNRAS.446.2380B,2014MNRAS.439.1160R,Wise2019}. 
With accretion at the Eddington limit, obtaining such masses during the available time is 
very difficult when starting from a stellar mass black hole \citep{Shapiro05}. 
Given the difficulties with other scenarios like photo-evaporation of HII regions around the 
first stars, the direct collapse is the most plausible scenario where a metal-free 
protogalactic halo collapses directly to form a 
massive BH \citep{Bromm2003,Begelman2006,Volonteri2006,Latif2013BHform,Wise2019}.

Molecular hydrogen is the only coolant in primordial haloes which can bring the temperature of 
the gas down to a few hundred Kelvin and may induce fragmentation by reducing the Jeans mass. The gas will show little fragmentation to form stars in zero metallicity haloes in the absence of H$_{2}$ cooling \citep{Li2003} which can be achieved via a strong photodissociating UV background in primordial haloes \citep{Dijkstra2008,Shang2010, Schleicher10}. In the presence of only atomic line cooling, the Jeans mass becomes about two orders of  larger (for $ n =10^{4}\,\mathrm{cm}^{-3}$ and $T=8000$\,K, $M_{J} \sim 10^{5}$\,\Msun) and fragmentation to lower mass scales remains inhibited. Numerical simulations show that the collapse of metal-free haloes with $ T_{vir} > 10^{4}$\,K  irradiated by strong LW flux proceeds isothermally in the absence of H$_{2}$ cooling and a massive object of $ 10^{5}-10^{6}$\,\Msun~can be formed  \citep{Bromm2003,Wise2008,Latif2014UV}.

Magnetic fields are expected to influence the formation of black holes by enhancing the Jeans mass due to the extra magnetic pressure and providing additional means for the transport of angular momentum by magnetic torques. The latter may become significant in the central accretion disc, implying the presence of strong rotation measures and enhanced accretion rates. 
In fact, the detection of strong rotation measures in quasars at z = 5.3 indicates the relevance of magnetic fields in the early universe \citep{Hammond2012}. The observations of nearby active galactic nuclei suggest that magnetic fields play a vital role in the transport of angular momentum \citep{Beck1999,Beck2005}.

Similarly, turbulence can affect the BH formation process by locally compressing the gas and regulating the transport of angular momentum. The hydrodynamical simulations employing a subgrid-scale (SGS) turbulence model to account for unresolved turbulence show that turbulence favours the formation of self-gravitating accretion discs, suppresses fragmentation
(due to an additional unresolved turbulent pressure), and that a turbulent viscosity term contributes to the transport of angular momentum \citep{Latif2013BHmass,Latif2013MassiveHalos}. While fragmentation occasionally occurred, depending on the properties of the host halo, the formation of massive objects still proceeded.

Turbulence also amplifies weak magnetic fields by converting turbulent energy into magnetic energy a process known as the small scale dynamo. Strongly amplified magnetic fields may reach equipartition with the kinetic energy and become dynamically important. \cite{Latif2014} have shown that strongly ($\mathcal{O} (10^0 - 10^1)$\,G) amplified magnetic fields provide a substantial amount of support on small scales, surpassing thermal pressure support by a factor of ${\sim}10$, and suppress fragmentation. However, these simulation lacked an MHD SGS model that treats the combined effect of unresolved turbulence and magnetic field dynamics.

Small scale physical processes are often not captured by numerical simulations due to the limited spatial resolution.
Large eddy simulations (LES) address this issue by incorporating unresolved small scale processes such as turbulent dissipation via an SGS model 
\citep{Sagaut2006,2009lesc.book.....G,lrca-2015-2}.
Simulations that treat turbulent dissipation implicitly by means of a shock 
capturing  numerical method are usually referred to as implicit large eddy 
simulations \citep[ILES, see, e.g.,][]{grinstein2007implicit}.
In addition to pure turbulent dissipation other small scale processes have been modelled by
an SGS model in the context of astrophysics.
These processes include turbulent deflagration in Type Ia supernovae \citep{Ropke_2007}, thermal and turbulent feedback from supernovae in isolated disc galaxies \citep{10.1093/mnras/stu1119}, or turbulent pressure support in the intergalactic medium \citep{Iapichino2011} and during black hole formation in the early universe\citep{Latif2013BHform}. 
All these examples focus on hydro and/or thermodynamic processes.
Including magnetic fields and associated small scale processes such as magnetic reconnection,
inverse (i.e., up-scale) cascades, and small scale dynamo action present challenges in 
SGS modelling \citep{Miesch2015}. 
The majority of MHD  SGS models have been derived from their incompressible hydrodynamic 
counterparts, and only few have been tested in the compressible MHD 
regime \citep{Miki2008,Chernyshov2014,Grete2015}.
\cite{Vlaykov2016a} and \cite{Grete2016a,Grete2017} have developed an SGS model that 
explicitly takes into account compressibility effects in MHD turbulence 
and demonstrated its applicability from the subsonic to the highly supersonic regime.
Therefore, it is most suitable for the dynamics of a Direct Collapse scenario.

In this study, we perform for the first time cosmological simulations that employ 
a subgrid-scale model for unresolved MHD turbulence and investigate its impact on 
the formation of Direct Collapse BHs.
We adopt the limit of a very strong radiation background, where molecular 
hydrogen is fully dissociated so that cooling is regulated only via atomic 
hydrogen lines. 
Such conditions tend to strongly suppress fragmentation \citep{Latif2013BHform}.
We study the statistical properties of 9 haloes with  different 
initial magnetic seed field strengths and the treatment of unresolved turbulence 
in the context of direct collapse scenario.
We find that fragmentation is intermittent and the run-away collapse is predominantly 
determined by
the microphysics of the atomic gas, while magnetic fields and turbulence play a secondary role.

The article is organised as follows. In Section~\ref{sec:method} we present our simulation setup including details about employed turbulence and chemical models. In Section~\ref{sec:results} we present our main results  and we summarise our conclusions in Section~\ref{sec:conclusions}.

\section{Method}
\label{sec:method}

\subsection{Simulation setup}
We conduct cosmological zoom-in simulations with \Enzo \citep{Enzo2013} which is an open source multi-physics code\footnote{See \url{http://enzo-project.org/}}.
To perform cosmological zoom-in simulations, we first ran dark matter (DM) only simulations with $1024^3$ DM particles to identify haloes and reran them  
including baryons and additional physics. We employed the adaptive mesh refinement (AMR) approach to add dynamical refinement during the collapse. 

All simulations start at a redshift of $z = 100$ based on the $\Lambda$CDM-model by taking cosmological parameters from Planck 2015 data \citep{Planck2015}.
The simulation domain is a cubic box with a side length of 1~Mpc/h and initial conditions are generated\footnote{Sample parameter files used for creating initial conditions
and running both types of cosmological simulations are available as supplementary 
online material.} with \Music \citep{Hahn2011}. 

We use the \Rockstar halo finder \citep{Behroozi2013} to identify the most massive haloes at $z~=~12$ and traced back all particles that were within 4 virial radii of that halo to their original positions at $z~=~100$ to recreate nested grid initial conditions\footnote{See a script by John Wise \textsc{get\_halo\_initial\_extent.py} at \url{https://bitbucket.org/jwise77/enzo-mrp-music}}. 
The newly generated initial conditions have a root grid with $256^3$ cells (and an equal number of DM particles) with two nested grids of the same resolution that enclose all previously identified particles. 
In each nested grid the spatial resolution doubles so that the effective spatial resolution of the region of interest is initially identical to the previous DM only run.
In total, we use 23,024,594 DM particles with an effective DM resolution of 99\,\Msun.

In order to follow the collapsing gas in the multi-physics simulations we use AMR that is triggered by one of the three following conditions: gas overdensity of a 
factor 4 (with a super-Lagrangian refinement exponent of $-0.3$), dark matter overdensity of a factor 4, and resolving the Jeans length by at least 64 
grid cells throughout the simulations. The latter is required in order to resolve turbulence  necessary for the small scale dynamo action 
\citep{Latif2013SSD}. DM particles are smoothed at level 12 to prevent artificial effects. We stop all simulations at a proper peak density of $10^{-9}$\,g/cm$^3$, 
corresponding to a number density of $\approx4.9\times10^{14}$\,cm$^{-3}$.
This is reached at a redshift of $z \approx 11.8$. At this final time the central region of the halo is represented on a grid at the 29th level of refinement corresponding to an effective spatial resolution of 0.17\,au (proper). 

We solve the cosmological magnetohydrodynamic equations using the MUSCL-Hancock framework with piecewise linear reconstruction and the HLL Riemann solver.
For increased numerical stability, we set \verb#Theta_Limiter = 1.0#, corresponding to a minmod flux limiter. The divergence constraint of the magnetic field is maintained by using hyperbolic divergence cleaning \citep{Dedner2002}. The chemical model employed in all simulation is described in more detail in 
subsection~\ref{sec:chem}.

In total, we conducted and analysed 36 simulations that vary with respect to initial conditions, initial magnetic field strength, and usage of a subgrid-scale (SGS) model for unresolved compressible MHD turbulence, see the following subsection~\ref{sec:sgs} for more details. 
The initial conditions were varied by using 9 different random seeds in \Music.
Given that our random seeds varied by 1 the random number
generator in \Music produced similar cosmological 
initial conditions\footnote{
See \url{https://groups.google.com/forum/\#!topic/cosmo\_music/BGrUohOlbsE} 
for a discussion.}.
Here, similar means that the large scales in the initial conditions are virtually identical
\begin{figure}
\centering
\includegraphics{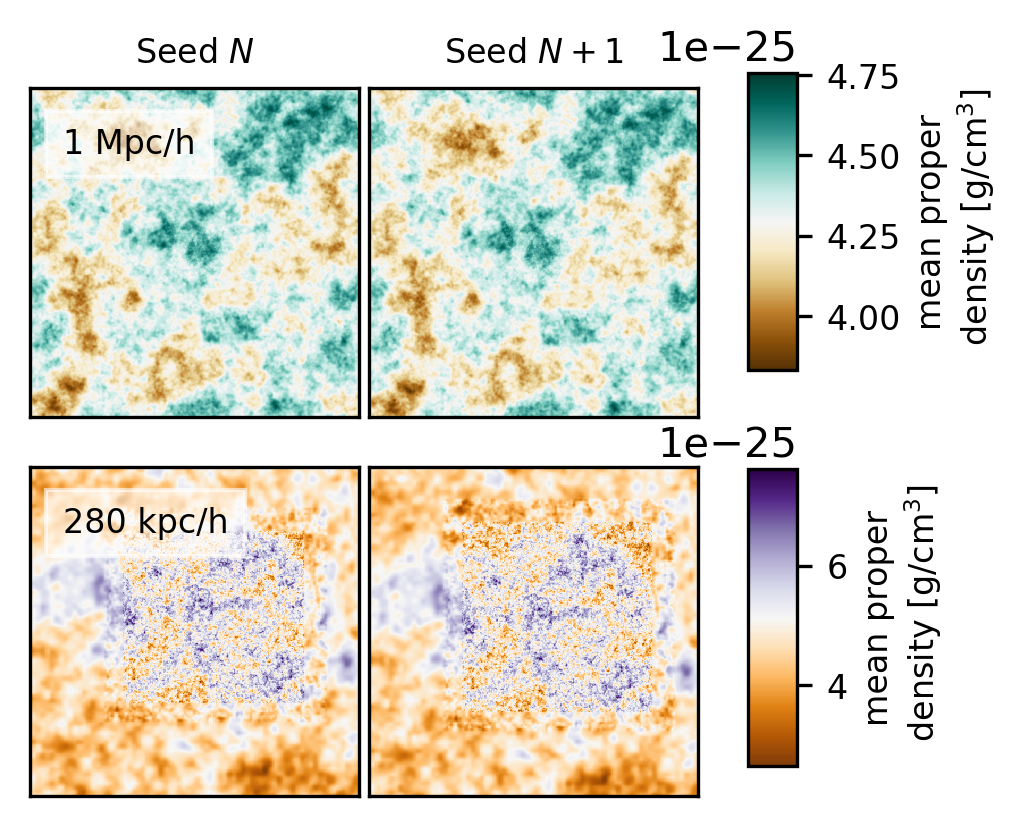}
\caption{Initial mean proper density along the z-axis of the entire domain (top panels) 
and proper density slice of 280\,kpc/h side length through the center of the box 
(bottom panels) for two different initial conditions.
The different spatial resolutions of level 1 and 2 of the
nested initial conditions are visible in the bottom panels.}
\label{fig:init-cond}
\end{figure}
as illustrated in Fig.~\ref{fig:init-cond} for two initial conditions whose seed varies by 1.
The root mean square of the relative density variation between those two initial conditions
over the entire domain is
\begin{eqnarray}
\sqrt{
\left < \left ( \frac{| \rho_1(\mathbf{x}) - \rho_2(\mathbf{x})|}
{0.5 \left(\rho_1(\mathbf{x}) + \rho_2(\mathbf{x})\right)} \right )^2 \right >
} = 0.09 \;.
\end{eqnarray}
Over the course of the evolution, the small variations in the initial conditions 
result in a sample of 9 similar haloes
(e.g., with respect to virial radii and masses, see Sec.~\ref{sec:results-overview})
that we use for statistical analysis.
The similarity between the 9 haloes allows us to explore the effect of very small variations in the initial conditions on the final small scale halo 
properties. In other words, we are able to differentiate between robust statistics, which are not sensitive to small initial variations, and statistically 
varying results that are prone to small physical variations and/or numerical effects.
All 9 haloes have been initialised with a proper uniform initial magnetic field strength of both $10^{-10}$\,G and $10^{-12}$\,G
corresponding to comoving magnetic field strengths of $10^{-5}$\,nG and $10^{-8}$\,nG, respectively. This is well below current estimates of the upper 
limit of, e.g., $4.4$\,nG (comoving at 1 Mpc scale) from the CMB power spectrum \citep{2016A&A...594A..19P}, see \citet{Subramanian2016} for a recent 
review.In addition, all 9 haloes were simulated with an explicit nonlinear SGS model (the LES case) and without an explicit model (the ILES case). We will refer 
to the simulations with different physics throughout the paper as ILES-B10, ILES-B12, LES-B10, and LES-B12, respectively, see Table~\ref{tab:frag} for a list of all simulations.

\subsection{Subgrid-scale turbulence model}
\label{sec:sgs}
To include the effect of unresolved, i.e., below the grid-scale, turbulence we employ the nonlinear SGS model for MHD turbulence presented by 
\citet{Vlaykov2016a,Grete2016a}. The main features of this model are that it specifically takes compressibility effects into account and allows for
energy transfer in both directions. In other words, energy can be transferred from large to small scales (e.g., the forward cascade and dissipative effects) 
and from small to large scales (e.g., the inverse cascade). 
The model has extensively been tested \textit{a priori} \citep{Grete2016a} and 
\textit{a posteriori} \citep{Grete2017} in different turbulent regimes ranging from the subsonic to
the highly supersonic regime.
Thus, it suitable for application in simulations of dynamic direct collapse scenarios.
The model is incorporated via two additional terms in the cosmological (comoving) MHD 
equations\footnote{See \citep{Enzo2013} for the general set of equations solved.}.

First, the turbulent stress tensor, 
\begin{eqnarray}
\tau_{ij} = \tu - \tb + \delta_{ij} \Emagsgs \;,
\end{eqnarray}
enters as a source term ($\ldots = -\frac{1}{a} \nabla \cdot \tau$) in the momentum equation
with $a$ as the cosmological scale factor.
It consists of the turbulent Reynolds stress, 
\begin{eqnarray}
\tu &=& 
\frac{1}{12} \Delta^2 \flt{\rho} \fav{u}_{i,k} \fav{u}_{j,k} \;, 
\end{eqnarray}
the turbulent Maxwell stress, 
\begin{eqnarray}
\tb &=& 
\frac{1}{12} \Delta^2  \flt{B}_{i,k} \flt{B}_{j,k} \;,  
\end{eqnarray}
and the turbulent magnetic pressure given via the turbulent magnetic energy $\Emagsgs$. In general, the turbulent kinetic and magnetic energies are 
given by the traces of the turbulent stresses via the identities
$2 \Ekinsgs = \tau_{ii}^{\mathrm{u}}$ and 
$2 \Emagsgs = \tau_{ii}^{\mathrm{b}}$.
Einstein summation applies and $\Box_{i,j}$ designates the $j$th partial derivative of component $i$ of $\Box$. The filtered density $\flt{\rho}$, 
the filtered velocity $\fav{u} = {\flt{\rho u}}/{\flt{\rho}}$, and the filtered magnetic field $\flt{B}$ are calculated via an explicit filter with
a filter width of $\Delta = 2.711 \Delta_x$ grid cells (see \citet{Grete2017} for details). The explicit filtering ensures a scale decomposition between 
the LES scale (on which the SGS terms are calculated) and the grid scale, which is most affected by numerical dissipation.

The second term of the model is the turbulent electromotive force, 
\begin{eqnarray}
\EMF &=&  
 \frac{1}{12} \Delta^2 \varepsilon_{ijk}  \bigl( \fav{u}_{j,l} \flt{B}_{k,l} 
 - \bra{\ln \flt{\rho}}_{,l} \fav{u}_{j,l} \flt{B}_{k} \bigr) 
\;,
\end{eqnarray}
which enters the induction equation as a source term ($\ldots~=~\frac{1}{a}\nabla\times\EMF$) with Levi-Civita symbol $\varepsilon_{ijk}$.

Finally, it should be noted that this SGS model falls in the category of instantaneous or zero-equation models. There is no additional dynamical equation
for the turbulent energies as the energies are determined instantaneously via the turbulent stresses. This also means that there is no intermediate 
reservoir of energy, as, for example is used in \citet{Schmidt2014} in the context of cosmological hydrodynamics.

\subsection{Chemical model} 
\label{sec:chem}
We use the chemical model described in \citet{Latif2016} and briefly summarise here its main features. We solve the chemical and thermal evolution of 
the gas using the \textsc{KROME} package \citep{Grassi2014}. In our simulations the rate equation of the following species, H, H$^-$, H$^+$, He, He$^+$, 
He$^{++}$, H$_2$, H$_2^+$, and e$^-$, are solved for non-equilibrium and are coupled to MHD. The chemical network and reaction rates are listed in
Appendix A of \citet{Latif2015}. It includes all the relevant processes for the formation and dissociation of H$_2$ as well as photo-detachment of H$^-$. 
We ignore the species involving deuterium as they get dissociated for even weaker radiation fluxes. In this work we assume a background Lyman Werner (LW) 
flux of strength $10^5$ in units of $J_{21}= 10^{-21}$\,erg/s/cm$^2$/Hz/sr with a fixed radiation temperature of $T_\mathrm{rad}= 2 \times 10^4$\,K. 
Such a choice of a strong LW flux ensures that collapse proceeds isothermally as it is well above the critical value of LW flux estimated from 3D 
simulations.

Furthermore, our model includes cooling due to the collisional excitation, collisional ionization, radiative recombination and Bremsstrahlung radiation. 
Also the cooling due to the molecular hydrogen, collisionally induced emission and chemical heating and cooling due to the three-body rates is included 
in our model. At high densities, we also take into account $H^-$ cooling as well as employ realistic opacities for bound-free $H^-$ and Lyman alpha 
cooling. For further details about our chemical model, see Section 2.2 of \citet{Latif2016}.

\section{Results}
\label{sec:results}
\subsection{Overview}
\label{sec:results-overview}
Overall, we observe a self-similar gravitational collapse 
that results in virtually identical large-scale properties of the most massive halo 
in all simulations independent of initial magnetic field strength and the unresolved
turbulence model.
This is expected as the large-scale properties of the initial conditions are practically
identical.
In general, the  most massive halo in every simulation, which we analyse in this paper,
has a virial radii of $\approx0.95$\,kpc and 
a mass of $\approx5.5\times10^{7}$\,\Msun (dark matter) and 
$\approx8.6\times10^{6}$\,\Msun (gas).
We stop all simulations at the same peak proper 
density of $10^{-9}$\,g/cm$^3$,
corresponding to a number density of $\approx4.9\times10^{14}$\,cm$^{-3}$.
The central region has an effective spatial resolution of 0.17\,au.
This allows us to differentiate between robust and statistically varying or 
intermittent properties on the small scales across haloes with different
initial magnetic field strength and unresolved turbulence model.

Given the overall similarities, we discuss the gas dynamics for only one representative halo (i.e., the set of four simulations with different physics for halo 8 selected arbitrarily) and also discuss the significant differences found in the complete suite of 36 simulations.

\subsection{Dynamical properties}
\begin{figure}
\centering
\includegraphics{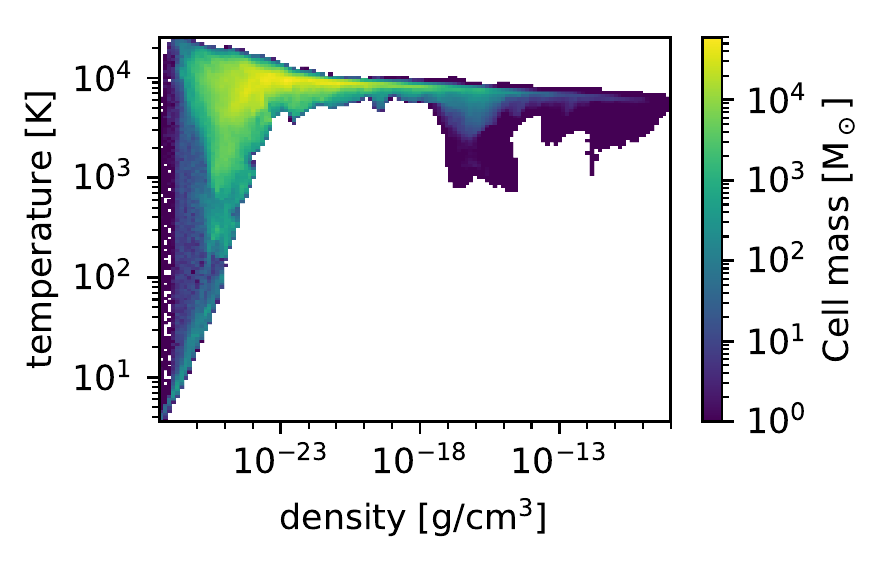}
\caption{Proper density-temperature phase diagram for a representative simulation (Halo 8 LES-B10).}
\label{fig:rho-T}
\end{figure}
\begin{figure*}
\centering
\includegraphics[width=\textwidth]{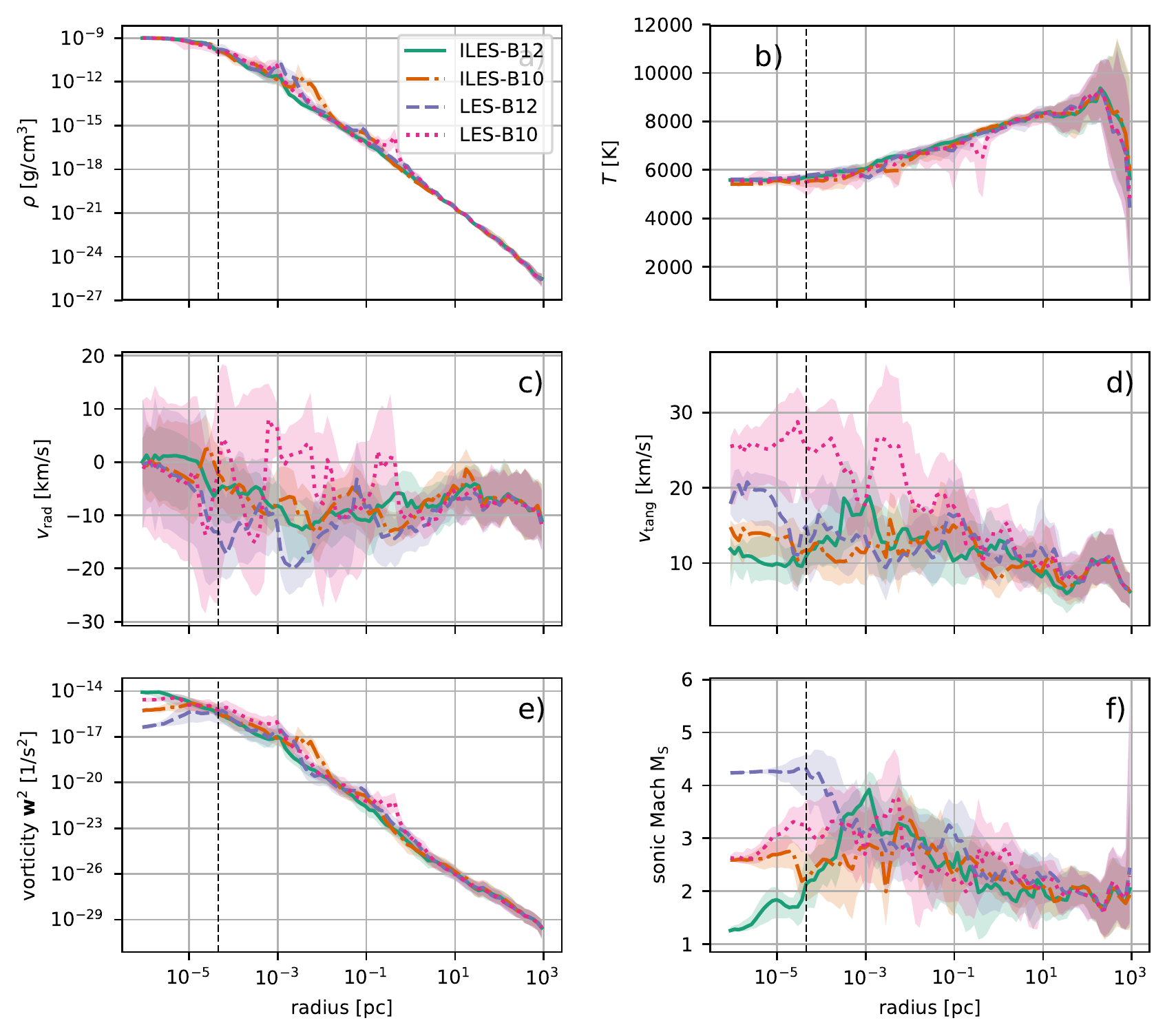}
\caption{Radial profiles of the median density (a), temperature (b), radial (c) and tangential (d) velocity, vorticity (e), and sonic Mach number (f),
for halo 8 at the same peak density of $10^{-9}$\,g/cm$^3$. Shaded areas indicate the interquartile range and the vertical dashed line the central Jeans length.
}
\label{fig:overview-prof}
\end{figure*}

\begin{figure*}
\centering
\includegraphics{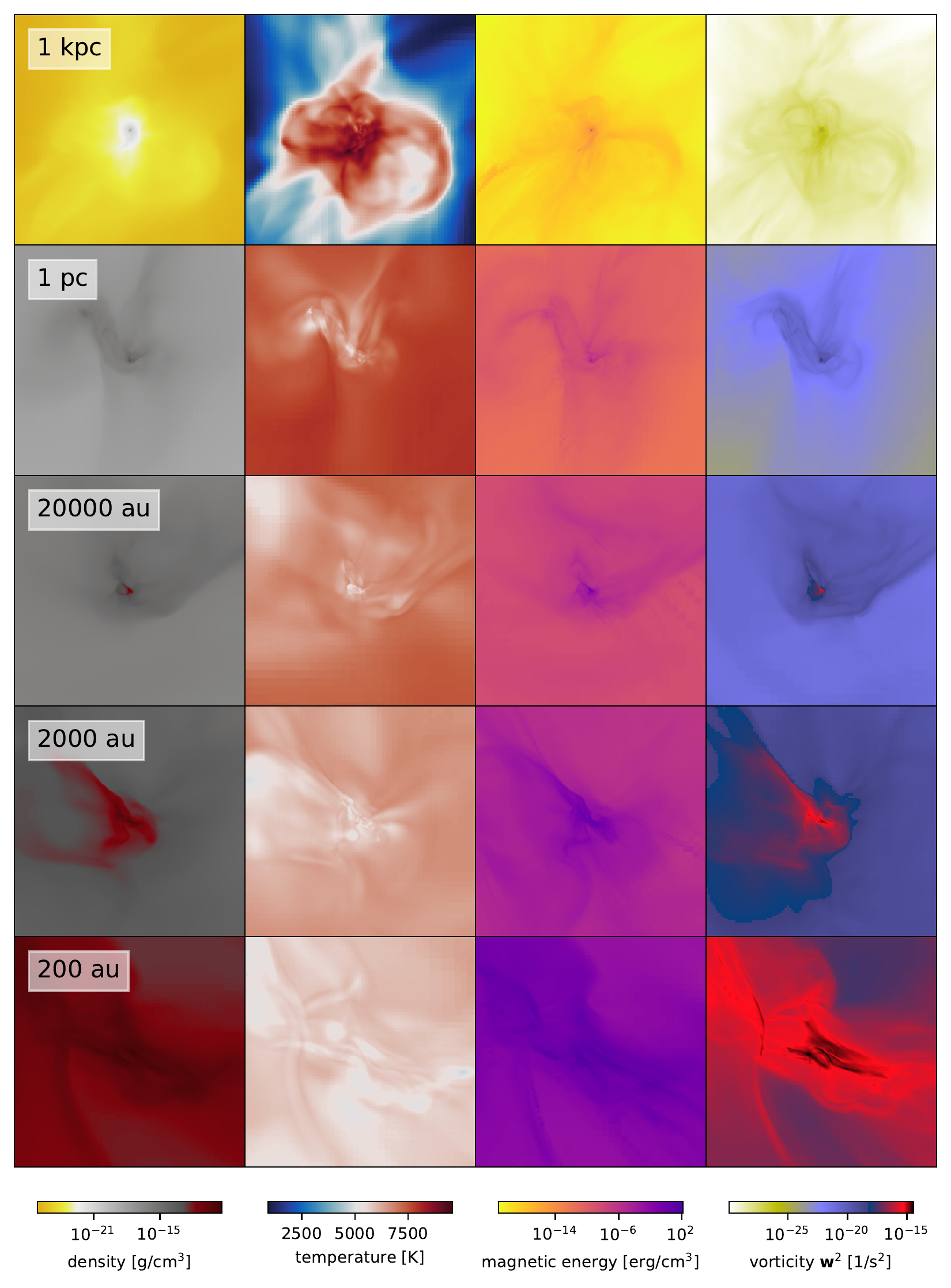}
\caption{Projections (density-weighted) of density, temperature, magnetic energy, and enstrophy (vorticity squared) centred on the peak density for the reference halo (Halo 8 LES-B10). The physical width shown in each panel is 1\,kpc 1\,pc, 20000\,au, 2000\,au, 
and 200\,au (from top to bottom row).
The two clumps of 3833\,\Msun and 1753\,\Msun identified by the clump finder
(see Table~\ref{tab:frag}) are visible in the 1\,pc panels.
}
\label{fig:overview-proj}
\end{figure*}

The dynamical properties of simulated haloes are discussed in this section. We show the density-temperature phase diagram for a representative halo in Fig.~\ref{fig:rho-T}. In the ubiquity of a strong LW radiation collapse proceeds isothermally with $T \approx~8000$\,K. At densities above $10^{-17}$\,g/cm$^3$ H$^{-}$ cooling becomes efficient and lowers the gas temperature down to $\approx~5500$\,K., see also the averaged temperature radial profile in Fig.~\ref{fig:overview-prof}(b).
This behaviour is consistent with previous studies investigating the impact of H$^-$ cooling \cite{Latif2016}.
The median density is shown in Fig.~\ref{fig:overview-prof}(a) and follows roughly $r^{-2}$ power-law outside the central Jeans length of $\approx10$\,au and 
gets flattened inside the Jeans radius. The small bumps on the density profiles are due to intermittent fragmentation (see also the density projection 
in the left column of Fig.~\ref{fig:overview-proj}).

Within the central region, the isothermal collapse stops (the infall velocity is $\approx0$\,km/s, 
see Fig.~\ref{fig:overview-prof}(c), and the density  profile is flattened) and further evolution will 
lead to the formation of an adiabatic core, see~\citet{Latif2016}. 
Outside of this region gas moves with a radial velocity of $\approx-10$\,km/s 
indicating large inflows towards the halo centre.
The tangential velocities are approximately constant throughout the inner parsec 
\begin{figure}
\centering
\includegraphics{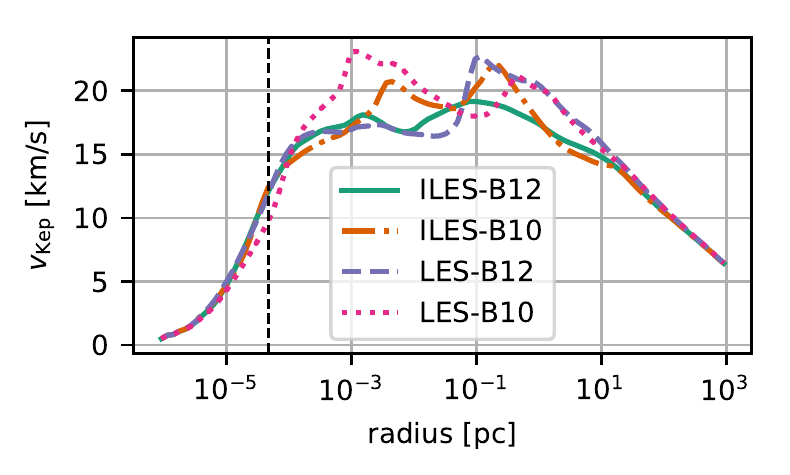}
\caption{Radial profiles of the Keplerian velocity for halo 8 at the 
same peak density of $10^{-9}$\,g/cm$^3$.}
\label{fig:orbvel}
\end{figure}
ranging between 10-20\,km/s
as depicted in Fig.~\ref{fig:overview-prof}(d) with the exception of LES-B10.
For comparison, the Keplerian velocities $v_\mathrm{Kep}~=~\sqrt{G M_\mathrm{encl}/r}$
are shown in Fig.~\ref{fig:orbvel}.
With 15-25\,km/s between 10\,au and 1\,pc the Keplerian velocities are generally 
equal to or larger than the tangential velocities.
In combination with infall velocities of $-10$\,km/s this suggests that there is little
support against collapse by coherent rotation in these haloes yet.
Similarly, no morphological changes from (turbulent) spherical distributions 
towards disk like structures are observed, see projections at various length scales 
for one halo in Fig.~\ref{fig:overview-proj}.
This equally applies to all other haloes in the sample.
Virtually all motion in the halo is supersonic 
with sonic Mach numbers ranging between $\Ms = 2-4$,
see radial profiles in Fig.~\ref{fig:overview-prof}(f).
Particularly the differences between ILES-B12 and LES-B12 in the inner region (e.g., $\Ms < 2$ for ILES-B12 and $\Ms > 4$ for LES-B12) may be attributed
 to the statistical variations, see Section~\ref{sec:variations}.

\subsection{Mass profiles and fragmentation}
\begin{figure}
\centering
\includegraphics[width=0.5\textwidth]{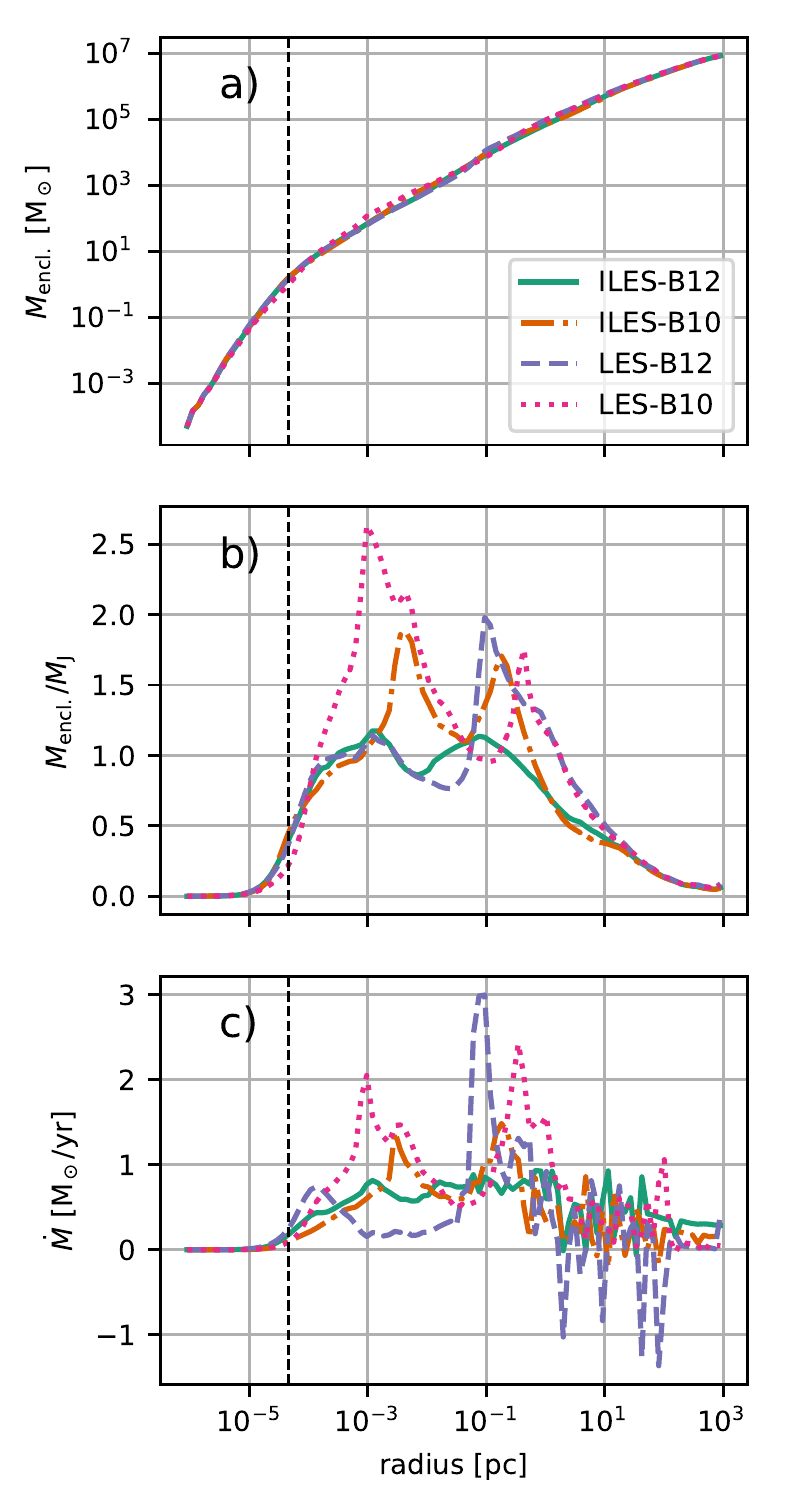}
\caption{Radial profiles of the enclosed mass (top), the ratio of the enclosed mass to the Jeans 
mass (centre) and the mass accretion rate (bottom) for halo 8 at the same peak density of $10^{-9}$\,g/cm$^3$.  The vertical dashed line indicates the central Jeans length.}
\label{fig:overview-mass}
\end{figure}

The enclosed baryon mass versus radius is illustrated in Fig.~\ref{fig:overview-mass}(a).
At the end of the simulations the central core within $\approx 10$\,au (proper) has a 
total mass a few \Msun.
The entire unstable core with $M_\mathrm{encl}/M_\mathrm{J} > 1$ extends to approximately
1\,pc with $M_\mathrm{J}~\approx~10^{5}$\Msun as shown in Fig.~\ref{fig:overview-mass}(b)
where
the ratio of the enclosed gas mass $M_\mathrm{encl}$ to the Jeans mass $M_\mathrm{J}$ 
versus radius is illustrated.
$M_{\mathrm{J}}$ is given by
\begin{eqnarray}
M_{\mathrm{J}} (r) = \frac{4\pi c_s^3(r)}{3 G^{3/2} \rho_0^{1/2}(r)} \,.
\end{eqnarray}
Here, $c_s(r)$ is the median sound speed at radius $r$ and 
$\rho_0(r)~=~3 M_\mathrm{encl}(r)/4\pi r^3$ the mean density.
For a gravitational collapse to occur this ratio $M_\mathrm{encl}/M_\mathrm{J}$
should be larger than one. Based on these estimates most of the gas within the central pc region is unstable to gravitational collapse.
The typical mass accretion rates shown in Fig.~\ref{fig:overview-mass}(c) are between $0.2-3$\,\Msun/yr.
The mass accretion rate at the edge of the unstable core is about 0.5\,\Msun/yr.
If this mass accretion rate persists for 1\,Myr, we expect the central core to reach 
$5\times10^5$ solar masses within the 1\,Myr.
Thus, they are well above the minimum value required
to form a direct collapse black hole \citep{2013A&A...558A..59S,2012ApJ...756...93H,2013ApJ...778..178H,2018MNRAS.474.2757H}.
The spikes in the mass accretion are due to the substructure in the halo and  no strong differences are observed between LES and ILES runs.

In order to  quantify the  fragmentation in our simulations we employ \yt's 
clump finder\footnote{
We consider density levels in factors of 2 and only track clumps that are gravitationally bound throughout the process.}
to find gravitationally bound clumps within the haloes at two different times: at a peak density of $10^{-12}$\,g/cm$^3$ and at a peak density of $10^{-9}$\,g/cm$^3$. The time difference between the two peak densities is $\approx 150$\,yr (for a reference,the free-fall time at 1\,pc is $\approx 50,000$\,yr).
\begin{table}
\centering
\begin{tabular}{lcccc}
\hline
		 & ILES-B12	 & ILES-B10	 & LES-B12	 & LES-B10	  \\\hline
\multirow{2}{*}{halo 1}		 & 281 \& 384 *		 & 1335 \& 1680 *		 & \multirow{2}{*}{no}	 & \multirow{2}{*}{no}\\
		 &  4040\,au		 & 23126\,au		 & 	 & \\
\multirow{2}{*}{halo 2}		 & \multirow{2}{*}{no}	 &  \multirow{2}{*}{no}	 & \multirow{2}{*}{no}	 & 831 \& 588	\\
      		 &   	 &    	 &   	 & 17908\,au \\
 
halo 3		 & no	 &  no	 & no	 & no\\
halo 4		 & no	 &  no	 & no	 & no\\
\multirow{2}{*}{halo 5}		 & \multirow{2}{*}{no}	 &  \multirow{2}{*}{no}	 & \multirow{2}{*}{no}	 & 356 \& 156 *	\\
      		 &   	 &    	 &   	 & 3227\,au \\
\multirow{2}{*}{halo 6}		 & \multirow{2}{*}{no}	 &  \multirow{2}{*}{no}	 & \multirow{2}{*}{no}	 & 2 \& 4 *	\\
      		 &   	 &    	 &   	 & 64\,au \\
\multirow{2}{*}{halo 7}		 & 5589 \& 5185	&  \multirow{2}{*}{no}	 & \multirow{2}{*}{no}	 & 	\multirow{2}{*}{no}\\
      		 & 56156\,au  	 &    	 &   	 &  \\
\multirow{2}{*}{halo 8}		 & \multirow{2}{*}{no}	 &  28 \& 42*		 & \multirow{2}{*}{no}	 & 3833 \& 1753	\\
      		 &   	 &  672\,au  	 &   	 & 76893\,au \\
\multirow{2}{*}{halo 9}		 & \multirow{2}{*}{no}	 &  \multirow{2}{*}{no}	 & 128 \& 121	& \multirow{2}{*}{no}\\
      		 &   	 &    & 3352\,au 	    	 &  \\
\hline
\end{tabular}
\caption{Fragmentation results of all 36 simulations, i.e.,  9 different haloes (with slightly different initial conditions) each with 4 different setups.
The masses of the clumps are given in \Msun~and the distance between their centers of mass is proper au. The asterisk indicates whether the fragmentation was first identified at a peak density of $10^{-9}$\,g/cm$^{3}$ and not yet at a peak density $10^{-12}$\,g/cm$^{3}$. The two snapshots are $\approx150$ years apart (compared to a free-fall time of $\approx 50,000$ years at 1\,pc).
 The fragments of halo 8 LES-B10 are visible in the 1\,pc projections of Fig.~\ref{fig:overview-proj}.
}
\label{tab:frag}
\end{table}

Table~\ref{tab:frag} lists the number of clumps and their masses in all simulations.
Only in 9 out of 36 simulations more than one clump is formed and in those cases usually there are  two clumps  of similar masses. 

However, these clumps are formed at small scales and are expected to merge within a dynamical time based on estimates of 
\cite{2015A&A...578A.118L}.  Overall, our results suggest that the process of fragmentation is stochastic and intermittent for our chosen setup and under the assumption of 
atomic hydrogen cooling. Given the large accretion rates 
and expected merging of clumps the formation of a central massive object is the most likely outcome. 

\subsection{Magnetic field dynamics}

\begin{figure}
\centering
\includegraphics[width=0.5\textwidth]{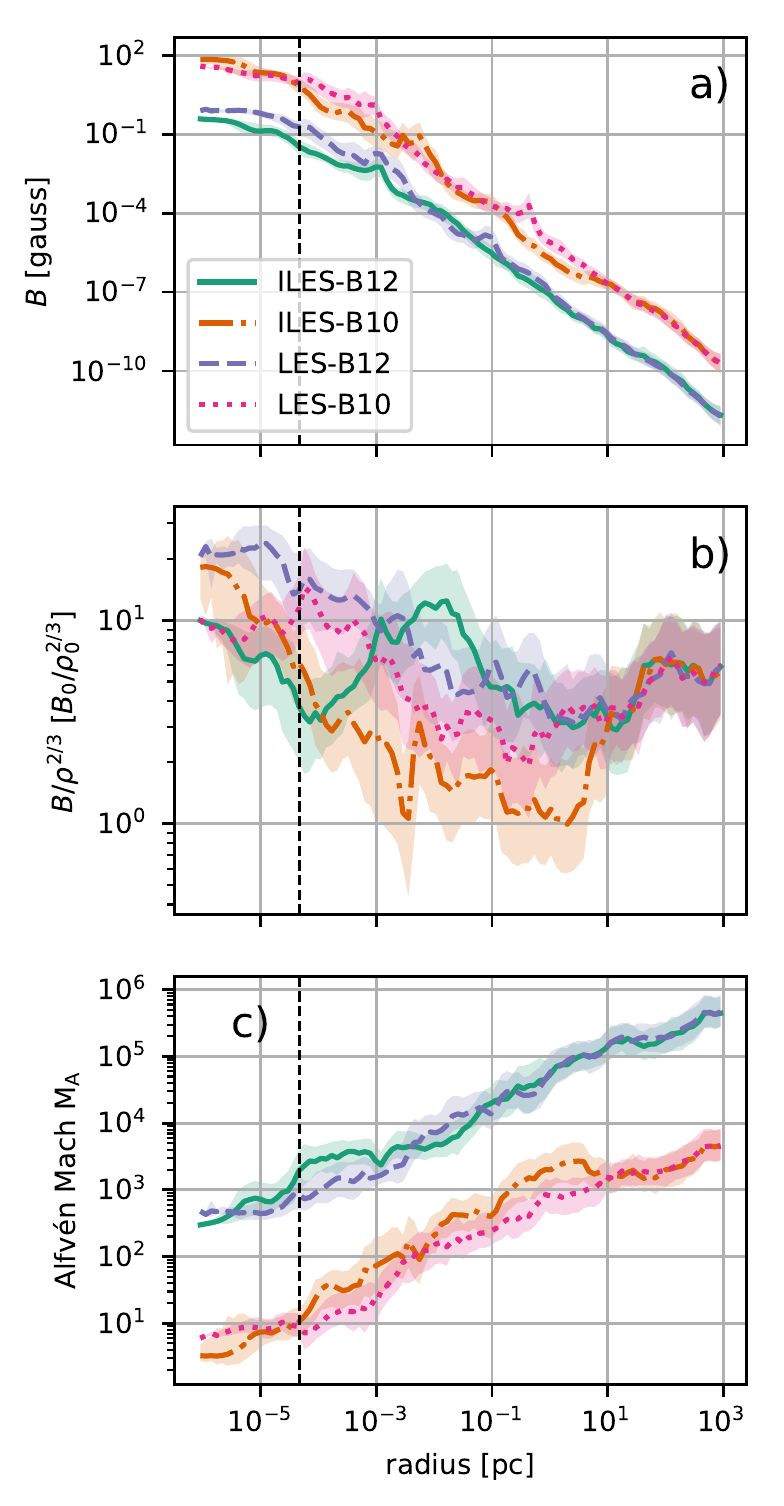}
\caption{Radial profiles of the magnetic field strength (a), the normalised magnetic field amplification (b), and the Alfv\'en Mach number (c)
for halo 8 at the same peak density of $10^{-9}$\,g/cm$^3$. The vertical dashed line indicates the central Jeans length.}
\label{fig:overview-B}
\end{figure}

The magnetic field strength increases towards the centre of the halo, as shown in Fig.~\ref{fig:overview-B}(a) for one reference halo.
No significant differences in magnetic field amplification are observed between LES and ILES runs. Simulations with larger initial seed fields 
have larger magnetic field strength but the overall amplification is the same for both cases.

To quantify the amplification of magnetic fields by the small scale dynamo, we have divided the magnetic field strength by $\rho^{2/3}$, the maximum possible 
contribution from flux freezing under spherical symmetry, and the result is shown in Fig.~\ref{fig:overview-B}(b).
At the 0.1-1\,pc scale the magnetic field amplification by the small scale 
dynamo action is a factor of a few and increases up  to a factor of 10 in the centre.
These results are in agreement with previous studies showing a critical 
resolution of at least 32-64 cells per Jeans length for dynamo action to occur \citep{2010ApJ...721L.134S,2011ApJ...731...62F,Latif2013SSD}.
In principle, we note that a larger amplification would be expected at even higher resolutions, 
as the Reynolds number of the flow is effectively limited by the numerical resolution.
Finally, for this particular halo the differences in the inner region (e.g., amplification of ${\lesssim}10$ for ILES-B12 and ${\sim}20$ for LES-B12) 
may indicate differences between the runs (e.g., between with and without SGS model), 
but our sample of different haloes allows us to attribute this to statistical variations as detailed 
in Section~\ref{sec:variations}.

The Alfv\'enic Mach number $\Ma$, i.e., the ratio of gas velocity to local Alfv\'en velocity $v_{\mathrm{A}} = B/\sqrt{\rho}$, is shown in Fig.~\ref{fig:overview-B}(c) and is a proxy for the ratio of kinetic to magnetic energy density. Kinetic motion clearly dominates the dynamics outside 
of the central region with $\Ma \approx 10^3$ (for ILES-B10 and LES-B10) and $\Ma \approx 10^4$ (for ILES-B12 and LES-B12) at $1$\,pc.
$\Ma$ decreases towards the centre in agreement with an active small-scale dynamo that converts kinetic energy to magnetic energy. For weaker initial seed 
fields the Alfv\'enic Mach number is 100 in the central region while for stronger initial fields the Alfv\'enic Mach number drops below 10 in the core but remains super-Alfv\'enic throughout. This suggests that even in the strong field case motions remain primarily driven by kinetic dynamics rather than magnetic field dynamics.

\subsection{Pressure support}
\begin{figure*}
\centering
\includegraphics[width=\textwidth]{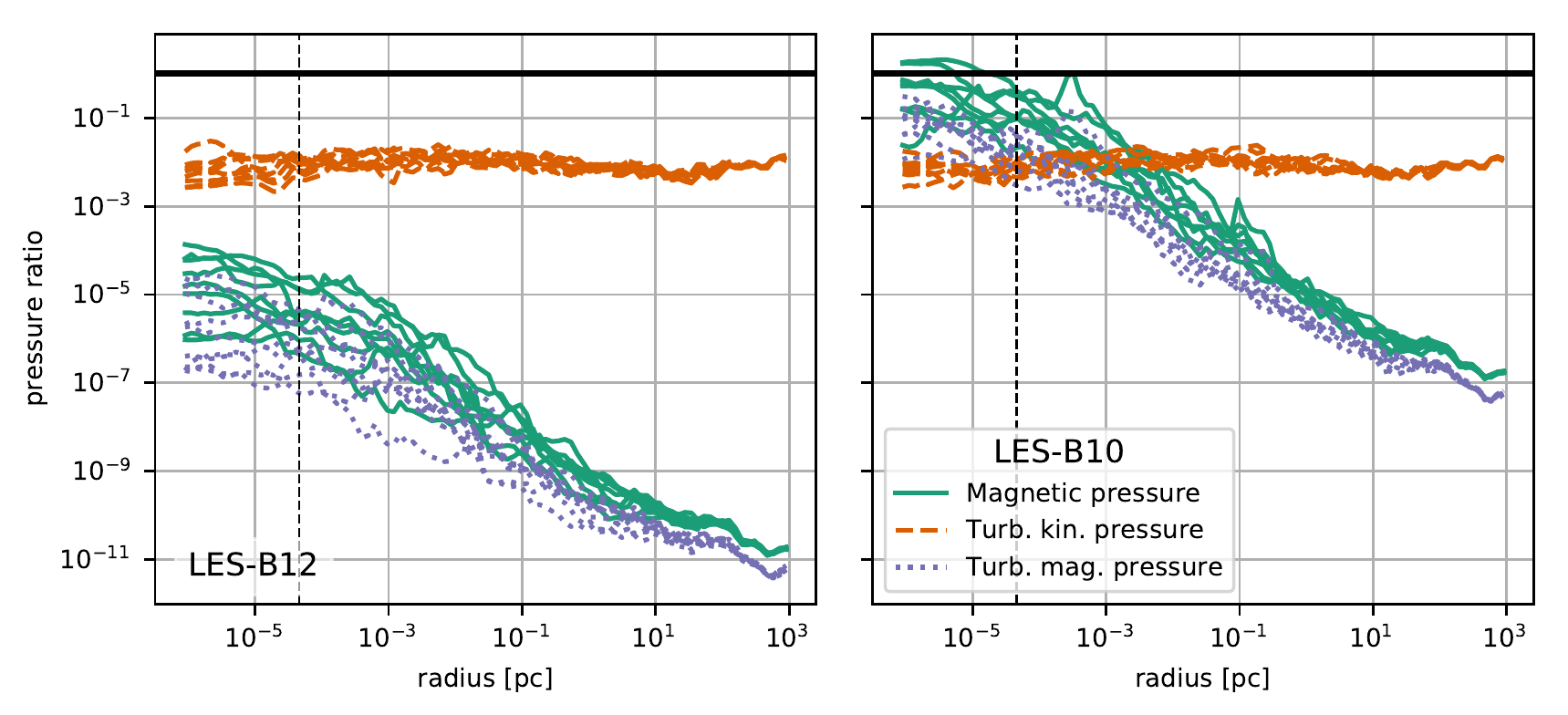}
\caption{Radial profiles of pressure ratios in LES. The green solid lines show the ratio of magnetic pressure to thermal pressure ($P$), 
the orange dashed lines show the ratio of turbulent kinetic pressure to $P$, and the dotted violet lines show the ratio of turbulent magnetic pressure 
to $P$. The horizontal black line at illustrates equilibrium with thermal pressure for guidance. The vertical dashed line indicates the central Jeans length.}
\label{fig:pres-balance}
\end{figure*}
Apart from thermal pressure, the presence of magnetic fields introduces an additional pressure, the 
so-called magnetic pressure, that may support 
the gas against fragmentation. Furthermore, in our LES two additional pressure terms, the turbulent kinetic and turbulent magnetic pressure,  are given by $\Ekinsgs$ and $\Emagsgs$, respectively. These terms take into account the cumulative effect of numerically unresolved velocity and magnetic field fluctuations. In the following, turbulent kinetic and magnetic pressure always refer to subgrid scales as opposed to the resolved kinetic and magnetic pressure, which encompasses all (turbulent and non-turbulent) contributions from scales larger than the grid resolution scale.  The ratio of the three non-ideal hydrodynamic pressures to the thermal pressure for all LES 
is illustrated in Fig.~\ref{fig:pres-balance}.

In the case of weak initial seed fields (LES-B12, left panel), both resolved and turbulent magnetic pressures are weak (<$10^{-4}$ relative 
to the thermal pressure).  For strong initial seed field cases  the resolved magnetic pressure reaches close to equipartition with the thermal pressure 
within the central core. This suggests that apart from some variations from halo to halo the presence of magnetic fields in the central region 
becomes significant. In all cases the turbulent magnetic pressure is approximately 10-20\% of the resolved magnetic pressure indicating that 
the overall magnetic field  dynamics is just barely resolved. The ratio of turbulent kinetic pressure to thermal pressure remains constant 
($\approx10^{-2}$) for both cases.

\subsection{Statistical variations}
\label{sec:variations}
\begin{figure*}
\centering
\includegraphics[width=\textwidth]{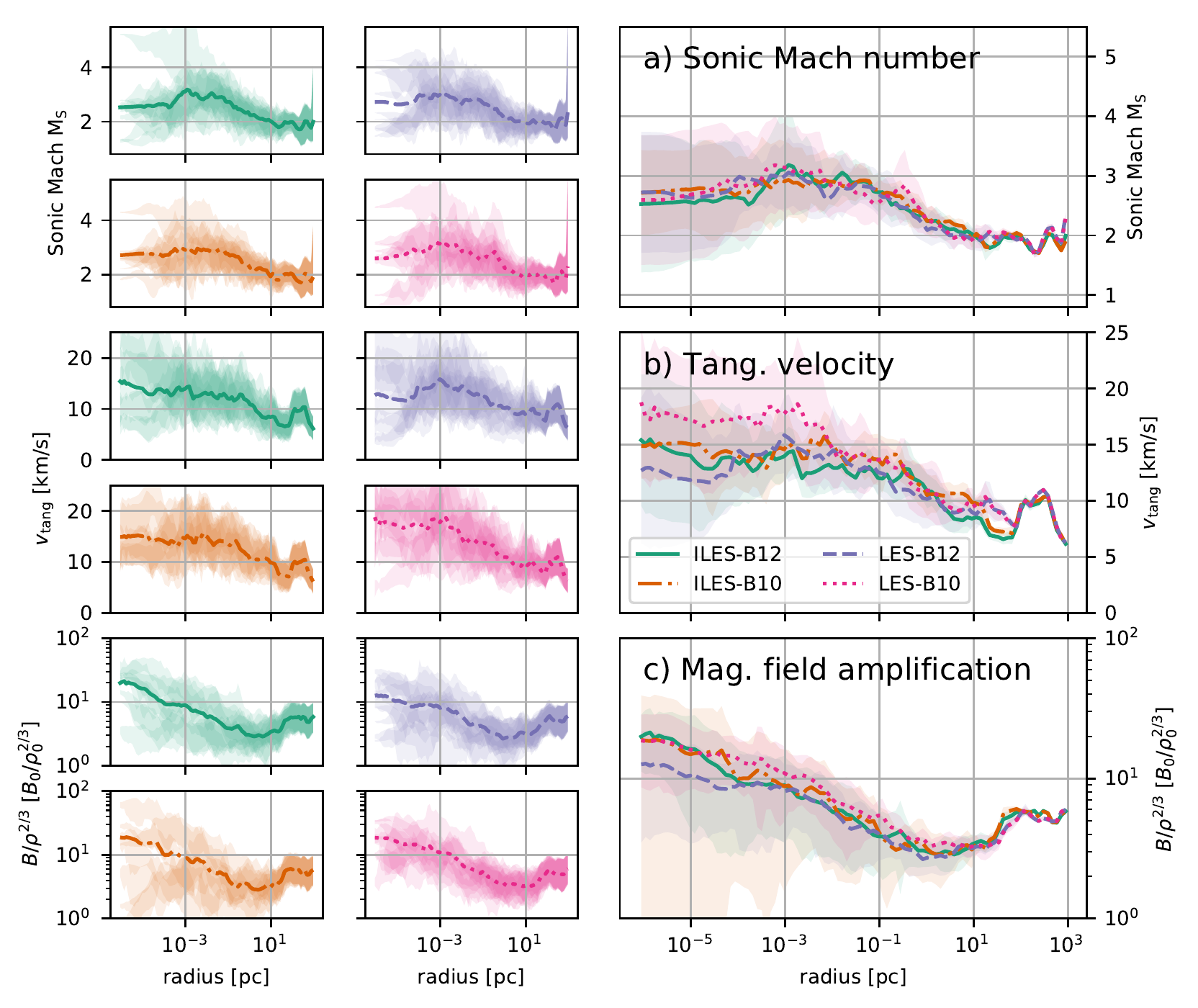}
\caption{Radial profiles of the sonic Mach number (a), tangential velocity (b), and magnetic field amplification (c) of all simulations at a peak density
of $10^{-9}$\,g/cm$^3$. The transparent regions on the left half correspond to the interquartile ranges of the individual simulations.
A region appears darker with increasing overlap of haloes. The individual lines correspond to the mean (over 9 haloes) of the median 
radial profile. For easier comparison between the configurations the mean lines are plotted together in the panels on the right including shaded regions 
that illustrate the standard deviation of the medians (between haloes).}
\label{fig:variations}
\end{figure*}

The majority of quantities discussed in previous sections are in agreement between  the 9 different haloes and across the 4 configurations.
A few exceptions, e.g., the tangential velocity $v_\mathrm{tang}$ in Fig.~\ref{fig:overview-prof}~(d), the sonic Mach number $\Ms$ in Fig.~\ref{fig:overview-prof}~(f), or the magnetic field amplification $B/\rho^{2/3}$  in Fig.~\ref{fig:overview-B}~(b), exhibit variations
between the configurations for the specific halo discussed before. Fig.~\ref{fig:variations} illustrates the variations in the radial profiles
of these quantities for all haloes and configurations in our sample. Overall, the variations in different haloes for each configuration individually 
is significant ($1 \lesssim \Ms \lesssim 5$, and between one and two orders of magnitude for $v_\mathrm{tang}$ and $B/\rho^{2/3}$, respectively) as shown in the left panels. The differences observed between configurations for an individual halo are negligible compared to the statistical variation of the entire sample of haloes. This is 
illustrated in the right panels of Fig.~\ref{fig:variations} where the mean and standard deviations of the median profiles (of all haloes) are drawn on
top of each other. All profiles (of different configurations) lie within the variations of each other. Thus, there is no significant statistical difference between the configurations. 

In general, all profiles shown including the ones discussed in previous sections are well
converged from 10\,pc outwards to the virial radius (0.95\,kpc).
This highlights the similarities between the large-scale properties of haloes resulting
from the similar initial conditions used in combination with the 
self-similar  behavior of isothermal collapse.
\section{Conclusions and discussion}
\label{sec:conclusions}
We performed 36 cosmological magnetohydrodynamic large eddy simulations (implicit and explicit)  with adaptive mesh refinement, and followed the collapse 
of 9 similar metal-free massive haloes ($\approx5.5 \times 10^7$\,\Msun~at $z \approx 11.8$) down to sub-au scales.All simulations employed a 
realistic chemical model for 9 species and assume a super-critical LW background flux from a nearby star-forming galaxy. In this study, we particularly explored the influence of magnetic fields and turbulence on the formation of supermassive black hole seeds within the direct collapse scenario. Thus, each of the 9 haloes was simulated using 4 different setups:  a weak and a strong initial magnetic field field, and with or without a subgrid-scale model for 
unresolved MHD turbulence.

We find that, by and large, \textit{the self-similar run-away gravitational collapse is the main driver in determining the majority of the halo properties}.
The presence of (explicitly modelled) turbulence or strongly amplified magnetic fields is secondary. For example, all simulations agree with an isothermal 
collapse at $\approx 8000$\,K exhibiting an $r^{-2}$ density profile.
Within the central parsec, all velocities are supersonic ($\Ms \approx 2-4$) with infall velocities of $\approx 10$\,km/s and tangential velocities 
of $\approx 15$\,km/s.
No significant coherent rotational support against the collapse is observed in any
simulation yet.
The resulting \textit{mass accretion rates vary between $0.2-3$\,\Msun/yr, and, thus, are large enough to support the direct collapse scenario}. 

Overall, fragmentation is quite intermittent in our simulations and appears as a  stochastic process 
as it occurred mildly in 9 out of 36 simulations. Even clumps 
forming in these cases are expected to merge within a dynamical time scale.

In all cases, we observe magnetic field amplification beyond simple compression of frozen-in magnetic fields, indicating small-scale dynamo action. 
\textit{In the strong magnetic field case the final magnetic fluxes reach dynamically relevant strengths}. Given that the seed field in our strong 
field case is still $>10^5$ times weaker than the upper limit from CMB power spectra \citep{2016A&A...594A..19P} a further increase of the initial 
field strength is still realistic. These stronger seed fields may be amplified to dynamically relevant strengths at earlier time during the collapse 
and may potentially affect gas dynamics.
We also note that our results on the magnetic field amplification correspond only to lower limits, 
as the Reynolds number in the simulations is limited by the numerical resolution, 
and also the analysis through the MHD SGS model shows that the MHD turbulence is currently under-resolved.

Systematic differences between ILES and LES are marginal in averaged radial profiles, although the turbulent magnetic pressure on the grid scale contributes about 20\% to the total magnetic 
pressure -- a significant fraction in the context of the fundamental assumption of explicit LES.
While it has been found that the Jeans length needs to be resolved by at least $32-64$  cells to observe the onset of small-scale dynamo action \citep{2010ApJ...721L.134S,2011ApJ...731...62F,Latif2013SSD},
our results suggest that in some cases even a higher resolution is required to fully resolve magnetic fields. Strongly amplified fields are expected to stabilise collapse and help in transferring angular momentum making the Direct Collapse scenario scenario more feasible.

Finally, our simulated sample of haloes is limited. 
All 9 different initial conditions are virtually identical on the large scales 
and, thus, result in a very similar large scale evolution of the simulations,
e.g., with respect to the virial properties of the halo during collapse.
This allowed us to study the intrinsic variations of a run-away collapse, cf., the 
butterfly effect in cosmological simulations \citep{Genel2018}, in one particular environment. 
We demonstrated that some quantities, such as the magnetic field amplification factor
within the core of the halo, vary by almost two orders of magnitude for a fixed configuration.
Moreover, the large intrinsic variation even for haloes with similar properties outshines statistical changes introduced by the LES approach.
Eventually, larger samples of similar haloes in combination with larger samples of
different haloes (or environments) are necessary, requiring a carefully chosen ensemble of 
cosmological simulations in order to make precise quantitative statements.

\section*{Acknowledgements}
The authors thank Brian W.~O'Shea for useful discussions.
PG acknowledges funding by NASA Astrophysics Theory Program
grant \#NNX15AP39G. ML thanks the funding from United Arab Emirates University via startup grant No 31S372.
DRGS thanks for funding through Fondecyt regular (project code 1161247), 
Conicyt PIA ACT172033, Conicyt Programa de Astronomia Fondo Quimal QUIMAL170001, 
Conicyt Programa de Astronomia Fondo ALMA 31160001 and the ``Concurso Proyectos 
Internacionales de Investigaci\'on, Convocatoria 2015'' (project code PII20150171).
The simulations were performed and analyzed with the HLRN-III
facilities of the North-German Supercomputing Alliance
under Grant No. \texttt{hhp00039}.
\Enzo and \yt are developed by a large number of independent researchers 
from numerous institutions around the world.
Their commitment to open science has helped make this work
possible.

%\bibliographystyle{mnras}
%\bibliography{references}

\end{document}